\documentclass[%
 reprint,
 amsmath,amssymb,
 aps,
]{revtex4-1}

\usepackage{graphicx}% Include figure files
\usepackage{dcolumn}% Align table columns on decimal point
\usepackage{bm}% bold math
\usepackage{amsfonts}

\begin{document}

\title{Thermostatistics of the polymeric ideal gas}

\author{M. A. Gorji} \email{m.gorji@stu.umz.ac.ir} \author{K.
Nozari}\email{knozari@umz.ac.ir} \affiliation{Department of Physics,
Faculty of Basic Sciences, University of Mazandaran, P.O. Box
47416-95447 Babolsar, Iran}\author{B.
Vakili}\email{b-vakili@iauc.ac.ir} \affiliation{Department of
Physics, Chalous Branch, IAU, P.O. Box 46615-397, Chalous, Iran}

\begin{abstract}
In this paper, we formulate statistical mechanics of the polymerized
systems in the semiclassical regime. On the corresponding polymeric
symplectic manifold, we set up a noncanonical coordinate system in
which all of the polymeric effects are summarized in the density of
states. Since we show that the polymeric effects only change the
number of microstates of a statistical system, working in this
coordinate is quite reasonable from the statistical point of view.
The results show that the number of microstates decreases due to
existence of an upper bound for the momentum of the test particles
in the polymer framework. We obtain a corresponding canonical
partition function by means of the deformed density of states. By
using the partition function, we study thermodynamics of the ideal
gas in the polymer framework and show that our results are in good
agreement with those that arise from the full quantum consideration
at high temperature, and they coincide with their usual counterpart
in the limit of low temperature. \\
\begin{description}
\item[PACS numbers]
04.60.Nc, 04.60.Pp, 05.20.-y, 05.30.-d
\item[Key Words]
Polymer quantization, Thermodynamics
\end{description}
\end{abstract}
\maketitle
\section{Introduction}
While general relativity (GR) improved our understanding about the
Universe, its shortages are revealed when it is utilized to describe
dynamics of the Universe in the standard big bang cosmology
\cite{IVP}. With the advent of theories such as inflationary
scenarios in order to solve the initial value problem and its later
success in explaining the origin of the large scale structures, this
idea was formed that classical GR may fail to describe properly our
Universe (at least in such high energy regimes) \cite{Inflation}.
The fact is that while GR is a classical theory in its original
formalism, the quantum effects significantly become important in the
very early Universe. It is plausible to expect that the initial
value problem will be naturally resolved when a quantum theory of
gravity is applied. Although it seems that a complete theory of
quantum gravity is not yet made, its main candidates such as string
theory and loop quantum gravity revealed some fundamental aspects of
the ultimate theory. For example the existence of a minimal
measurable length of the order of the Planck length is a common
feature of any quantum theory of gravity \cite{String,Loop}.
Assuming a minimal invariant length, the Heisenberg uncertainty
principle trivially implies an ultraviolet cutoff for the system
under consideration. However, the standard uncertainty principle
cannot support existence of a minimal length and the standard
Schr\"{o}dinger representation of the quantum mechanics is no longer
applicable. Therefore, some attempts have been done to include a
fundamental length scale in the standard quantum mechanics, see for
instance \cite{GUP} in which the generalized uncertainty principle
is introduced with the existence of a minimum measurable length in
its formalism. The Hilbert space representation of such modified
uncertainty relation is formulated in Ref. \cite{GUP-HS}. In more
recent times, polymer representation of the quantum mechanics has
been studied in the context of loop quantum gravity \cite{QPR}. The
associated Hilbert space supports the existence of a minimal length,
here known as polymer length scale \cite{QPR2}. The relation between
polymer picture and the generalized uncertainty principle framework
is investigated in Ref. \cite{PLMR-GUP}. A notable character of the
polymer representation of quantum mechanics is that, in contrast to
the standard Schr\"{o}dinger representation, in the classical limit
$\hbar\rightarrow\,0$ the system does not tend to its usual
classical version, instead, one recovers a one-parameter
$\lambda$-dependent classical theory, where $\lambda \sim l_{\rm
poly}/\hbar$ denotes the polymer length scale. The standard
classical theory emerges in the continuum limit
$\lambda\rightarrow\,0$. Exploring the implication of the effective
classical $\lambda$-dependent theory gives some interesting results.
For instance, existence of an upper bound for the energy of a
classical systems and removing the big bang singularity in the
cosmological setup when it applies to the minisuperspace models
\cite{QPR2,CPR}. As another feature, we will see that the effective
$\lambda$-dependent classical theory reproduces some results of the
so-called doubly special relativity \cite{DSR} (in which a minimal
observer independent length scale is proposed to special
relativity), and polymer length plays the role of the observer
independent length scale.

The existence of an invariant minimal length has also interesting
effects on the thermodynamical behavior of the physical systems. In
this regard, many efforts have been made to formulate the
statistical mechanics in the framework of the generalized
uncertainty principle \cite{GUP-THR}. Also, thermodynamics of black
hole systems in the polymer picture is studied in
\cite{Acosta,PLMR-THR} and the statistical mechanics of the ideal
gas in doubly special relativity is investigated in \cite{DSR-THR}.
Nevertheless, in order to study the thermodynamics of a given
physical system, one needs to know the microphysics of the system
which in turn is determined by quantum mechanics. However, for
instance, finding the energy eigenvalues is not an easy task at all
when one takes into account the minimal length considerations in the
problem at hand \cite{GUP-HS,QPR3,Acosta}. In an alternative
picture, one can work with the Hamiltonian and density of states in
the corresponding phase space in the semiclassical regime. In this
paper, at first we try to formulate the classical phase space of a
polymerized systems in terms of the language of the symplectic
geometry. Then we formulate statistical mechanics of the polymerized
systems in the semiclassical regime. We show that our results are in
good agreement with their quantum counterparts.

The structure of the paper is as follows: In section 2, we define a
polymeric structure on the symplectic manifold. In section 3, we
obtain a deformed density of states that contains all the polymeric
effects. Using the deformed density of states, we find a canonical
partition function for the polymerized systems. In section 4, we
study the polymeric effects on the thermodynamics of the ideal gas
through the polymeric partition function. Section 5 is devoted to
the conclusions.
\section{Polymerization}
In the Hamiltonian formulation of the classical mechanics, the
kinematics of the phase space is formed by the Poisson algebra
\begin{equation}\label{CPA}
\{q,p\}=1\,,
\end{equation}
where $(q,p)$ are the phase space variables known as canonical
variables. Then, quantization is the passage from the classical
Poisson algebra (\ref{CPA}) to the quantum Heisenberg algebra by the
standard rule: Replacing the Poisson bracket by the Dirac commutator
for the operators counterpart of the canonical variables as
\begin{equation}\label{CCR}
[\hat{q},\hat{p}]=i\hbar\,\hat{\bf 1}\,,
\end{equation}
where $\hbar$ is the Planck constant. It is easy to see that the
Heisenberg uncertainty principle is a straightforward result of the
commutation relation (\ref{CCR}). The ordinary Schr\"{o}dinger
picture of quantum mechanics is based on the representation of
operators on the Hilbert space ${\mathcal H}=L^2({\mathbb R},dq)$,
the space of the square integrable functions with respect to the
Lebesgue measure $dq$ on the real line ${\mathbb R}$. In addition to
this well-known representation, there are other representations
based on which one can construct the quantum kinematics. Here, we
are going to pursue a case that has been presented in \cite{QPR2}
under the name of polymer representation. The polymer representation
of quantum mechanics is formulated on the Hilbert space ${\mathcal
H}_{\rm poly}=L^2({\mathbb R}_{_d},d\mu_{_d})$, where $d\mu_{_d}$ is
the Haar measure, and ${\mathbb R}_{_d}$ is the real line but now
endowed with the discrete topology \cite{PS-Topology}. The extra
structure in polymer picture is properly described by a
dimension-full parameter $\lambda$ such that the standard
Schr\"{o}dinger representation will be recovered in the continuum
limit $\lambda\rightarrow\,0$ \cite{QPR2}. Evidently, the classical
limit of the polymer representation $\hbar\rightarrow\,0$, does not
yield to the classical theory from which one has started but to an
effective $\lambda$-dependent classical theory which may be
interpreted as a classical discrete theory. Such an effective theory
can also be extracted directly from the standard classical theory
(without any attribution to the polymer quantum picture) by using
the Weyl operator \cite{CPR}. The process is known as {\it
polymerization} with which we will deal in the rest of this paper.

In polymer representation of quantum mechanics, the position space
(with coordinate $q$) is assumed to be discrete with discreteness
parameter $\lambda$ and consequently the associated momentum
operator $\hat{p}$, that would be a generator of the displacement,
does not exist \cite{QPR}. However, the Weyl exponential operator
(shift operator) correspond to the discrete translation along $q$ is
well defined and effectively plays the role of momentum for the
system under consideration \cite{QPR2}. Taking this fact into
account, one can utilize the Weyl operator to find an effective
momentum in the semiclassical regime. Therefore, the derivative of
the state $f(q)$ with respect to the discrete position $q$ can be
approximated by means of the Weyl operator as \cite{CPR}
\begin{eqnarray}\label{FWD}
\partial_{q}f(q)\approx\frac{1}{2\lambda}[f(q+\lambda)-f(q-
\lambda)]\hspace{2cm}\nonumber\\=\frac{1}{2\lambda}\Big(
\widehat{e^{ip\lambda}}-\widehat{e^{-ip\lambda}}\Big)\,f(q)=
\frac{i}{\lambda}\widehat{\sin(\lambda p)}\,f(q),
\end{eqnarray}
and similarly the second derivative approximation gives
\begin{eqnarray}\label{SWD}
\partial_{q}^2f(q)\approx\frac{1}{\lambda^2}[f(q+\lambda)-2
f(q)+f(q-\lambda)]\hspace{1cm}\nonumber\\=\frac{2}{\lambda^2}
(\widehat{\cos(\lambda p)}-1)\,f(q).\hspace{2cm}
\end{eqnarray}
Inspired by the above approximations, the polymerization process is defined for the finite values of the parameter $\lambda$ as
\begin{eqnarray}\label{Polymerization}
\hat{p}\rightarrow\,\frac{1}{\lambda}\widehat{\sin(\lambda p)},
\hspace{1cm}\hat{p}^2\rightarrow\,\frac{2}{\lambda^2}(1-
\widehat{\cos(\lambda p)}).
\end{eqnarray}
The (quantum) polymerization (\ref{Polymerization}) suggests the
classical polymer transformation ${\mathcal{P}}[F]$ of a function
$F(q,p)$ on the phase space as \cite{CPR}
\begin{eqnarray}\label{PT}
{\mathcal{P}}[F(q)]=F(q),\hspace{1.5cm}{\mathcal{P}}[p]=\frac{
1}{\lambda}\,\sin(\lambda p),\nonumber\\{\mathcal{P}}[p^2]=
\frac{2}{\lambda^2}(1-\cos(\lambda p)),\hspace{1.5cm}
\end{eqnarray}
and in a same manner one can find polymer transformation of the
higher powers of momentum $p$. In this sense, by a classical {\it
polymerized} system, we mean a system that the transformation
(\ref{PT}) is applied to its Hamiltonian.

Now, consider a nonrelativistic physical system with standard
Hamiltonian
\begin{equation}\label{Hamiltonian}
H=\frac{p^2}{2m}+U(q)\,,
\end{equation}
where $m$ is the mass of a particle moving under the act of the
potential function $U(q)$. Applying the polymer transformation
(\ref{PT}), the associated effective Hamiltonian will be
\begin{equation}\label{PHamiltonian}
H_{\lambda}=\frac{1}{m\lambda^2}\big(1-\cos(\lambda p)\big)
+U(q).
\end{equation}
The first consequence of the polymerization (\ref{PT}), which is
also clear from the Hamiltonian (\ref{PHamiltonian}), is that the
momentum is periodic and its range should be bounded as
$p\in[-\frac{\pi}{\lambda},+\frac{\pi}{ \lambda})$. In the limit
$\lambda\rightarrow\,0$, the effective Hamiltonian
(\ref{PHamiltonian}) reduces to the standard one (\ref{Hamiltonian})
and one recovers the usual range for the canonical momentum
$p\in(-\infty,+\infty)$. Therefore, the polymerized momentum is
compactified and topology of the momentum sector of the phase space
is $S^1$ rather than the usual $\mathbb{R}$. As we will see, this
structure for the topology of the phase space gives nontrivial
results for the polymeric thermodynamical systems, for instance,
existence of an upper bound for the internal energy of the system.
In order to study the thermodynamics of the polymerized systems, we
implement the symplectic geometry which gives a more suitable
picture from the statistical point of view.

\subsection{Darboux chart}
Consider a two-dimensional symplectic manifold ${\mathcal M}$
thought as a polymeric phase space with symplectic structure
$\omega$ which is a closed nondegenerate $2$-form on
${\mathcal{M}}$. According to the Darboux theorem \cite{Arnold},
there is always a local chart in which this $2$-form takes the
canonical form
\begin{eqnarray}\label{Dar-twoform}
\omega=dq\wedge\,dp,\,
\end{eqnarray}
and as we will see the variables $(q,p)$ may be identified with the
canonical variables which we have perviously defined in (\ref{CPA}).
Although the symplectic structure (\ref{Dar-twoform}) is canonical
in polymeric phase space, one should be careful about the periodic
condition for the canonical momentum which significantly leads to
the nontrivial topology for the momentum part of the manifold
${\mathcal M}$. Time evolution of the system is given by the
Hamiltonian vector field ${\bf x}_{_H}$ which satisfies the equation
\begin{equation}\label{VF-D}
i_{\bf x}\,\omega=dH_{\lambda}\,,
\end{equation}
where $H_{\lambda}$ is given by relation (\ref{PHamiltonian}). By
solving the above equation with the use of the effective polymeric
Hamiltonian given in (\ref{PHamiltonian}), one gets
\begin{eqnarray}\label{Dar-VF}
{\bf x}_{_H}=\frac{\sin(\lambda p)}{m\lambda}\frac{\partial}{
\partial q}-\frac{\partial U}{\partial q}\frac{\partial}{
\partial p}.
\end{eqnarray}
The integral curves of the Hamiltonian vector field (\ref{Dar-VF})
are the polymer-modified Hamiltonian equations of motion in
canonical chart
\begin{equation}\label{Dar-PHE}
\frac{dq}{dt}=\frac{\sin(\lambda p)}{m\lambda},\hspace{1cm}
\frac{dp}{dt}=-\frac{\partial U}{\partial q},
\end{equation}
which clearly reduce to the standard Hamilton's equations in the continuum limit $\lambda\rightarrow\,0$.

The Poisson bracket between two real valued functions $F$ and $G$ on
${\mathcal M}$ is defined as
\begin{equation}\label{PB}
\{F,\,G\}=\omega({\bf x}_{_F},{\bf x}_{_G})\,.
\end{equation}
The closure of the $2$-form ensures that the Jacobi identity is satisfied by the resultant Poisson brackets in definition (\ref{PB}).
In the Darboux chart, we have ${\bf x}_{_F}=\frac{\partial F}{\partial p}\frac{\partial}{\partial q}-\frac{\partial F}{
\partial q}\frac{\partial}{\partial p},$ and also the same expression for the function $G$. Substituting these results together with
the canonical structure (\ref{Dar-twoform}) into the definition (\ref{PB}), gives
\begin{eqnarray}\label{Dar-PB}
\{F,\,G\}=\frac{\partial F}{\partial q}\frac{\partial G}{\partial p}-
\frac{\partial F}{\partial p}\frac{\partial G}{\partial q},
\end{eqnarray}
which is obviously the standard canonical Poisson bracket between
two arbitrary functions $F$ and $G$. It is clear that with choosing
$F(q,p)=q$ and $G(q,p)=p$ we have
\begin{equation}\label{Dar-PA}
\{q,\,p\}=1\,,
\end{equation}
which is nothing but the standard canonical Poisson algebra
(\ref{CPA}). It is important to note that while the polymeric phase
space and the standard classical relation (\ref{CPA}) from which we
have started, seem to have the same kinematical structure,
dynamically they are different thanks to the polymer-modified
Hamiltonian equations of motion (\ref{Dar-PHE}). So, the Poisson
algebra is canonical in the Darboux chart, but the effective
Hamiltonian (\ref{PHamiltonian}) contains the polymeric effects
which will change the dynamics through the relation (\ref{VF-D}).

Now, suppose that the system under consideration is a many-particle
one for which we are going to apply the above formalism in
statistical mechanics point of view. To do this, it is necessary to
consider the well-known Liouville theorem which is directly related
to the number of accessible microstates of the system. The Liouville
volume for a $2D$-dimensional symplectic manifold is defined as
\begin{eqnarray}\label{Vol-D}
\omega^{_D}=\frac{(-1)^{D(D-1)/2}}{D!}\,\underbrace{\omega\,\wedge
\,...\,\wedge\,\omega}_{D\,\,\mbox{times}}.
\end{eqnarray}
As a special case we see that for two-dimensional manifolds, the
Liouville volume coincides with the symplectic $2$-form. According
to the Liouville theorem, the Liouville volume (\ref{Vol-D}) is
conserved along the Hamiltonian flow ${\bf x}_{_H}$ as
\begin{eqnarray}\label{L-T}
{\mathcal L}_{\bf x}\omega^{_D}=0,
\end{eqnarray}
where ${\mathcal L}_{\bf x}$ denotes the Lie derivative along the
Hamiltonian vector field ${\bf x}_{_H}$. This relation can be easily
deduced from (\ref{VF-D}) in which we have also noticed that
$d\omega=0$ due to the Cartan formula as ${\mathcal L}_{\bf
x}\omega=i_{\bf x}d\omega+di_{\bf x}\omega=0$ \cite{LT,LT2}. So, the
Liouville theorem is always satisfied on a symplectic manifold
independent of a chart in which the physical system is considered.

If we restrict ourselves to a finite one-dimensional spatial volume
$L$, the total volume of the phase space can be obtained by
integrating the Liouville volume as
\begin{eqnarray}\label{Dar-TV}
\mbox{Vol}(\omega^1)=\int\omega^1=\int_{L}dq\times\int_{
-\frac{\pi}{\lambda}}^{+\frac{\pi}{\lambda}}dp=2\pi\Big(
\frac{L}{\lambda}\Big).
\end{eqnarray}
In the standard Hamiltonian formalism of the classical mechanics,
the total volume of the phase space (\ref{Dar-TV}) diverges even if
one confines the physical system to a finite spatial volume. More
precisely, the momentum part of the integral of Liouville volume
will diverge because there is no any priori restriction on the
momentum of the test particles in the standard classical mechanics.
However, due to existence of an upper bound for the momenta in the
classical polymeric systems the resultant total volume
(\ref{Dar-TV}) will be naturally finite. In other words, compact
topology of the momentum part of the polymeric symplectic manifold
implies the finite value for the associated total volume that is
circumference of a circle with radius $\lambda^{-1}$ (see Ref.
\cite{LT} for more details).

Another point here is that, the spatial sector of the phase space
volume $L$ should be quantized with respect to the polymer length
$l_{\rm poly}=\alpha\hbar\lambda$, since the polymer length is the
possible minimum length for the polymerized system. Here $\alpha=
{\mathcal O}(1)$ is a dimensionless coefficient which should be
fixed by the experiments \cite{QGExperiment}. Therefore, we have
$\frac{L}{l_{\rm poly}}\in{\mathbb N}$. This result in some sense is
similar to the result obtained in Ref. \cite{GUP-D} in the
generalized uncertainty principle framework. By taking this fact
into account, relation (\ref{Dar-TV}) may be rewritten as
\begin{eqnarray}\label{Dar-QTV}
\mbox{Vol}(\omega^1)=nh,
\end{eqnarray}
where $n$ is a positive integer which counts the number of
fundamental cells $\hbar\lambda$ exist in $L$. This equation shows
that the total volume of the phase space is naturally quantized with
respect to the Planck constant $h=2\pi\hbar$. Note that in the
semiclassical regime, to obtain a finite number of microstates for a
given statistical system, one needs an extra assumption that the
volume of the phase space is quantized with respect to the Planck
constant. However, as our above analysis shows, this issue
automatically emerges in the polymerized phase space. The origin of
this result may be sought in the Heisenberg uncertainty principle.
In the standard phase space there is no restriction according to
which the system fails to have access to any desired length scale
even the zero length. However, in the polymer picture the theory is
equipped with a maximal momentum correspond to the polymer length
(below which no other length can be observed) in the light of the
uncertainty principle as $p_{\rm max}\sim\,\frac{\hbar}{l_{\rm
poly}}\sim {\lambda}^{-1}$, the existence of which is responsible
for the quantized volume of the phase space.

\subsection{Noncanonical chart}
In order to study the statistical physics of a polymerized systems,
we introduce a noncanonical transformation
\begin{equation}\label{NC-Transformation}
(q,p)\rightarrow\,(q',p')=\Big(q,\frac{2}{\lambda}\sin(\frac{
\lambda p}{2})\Big),
\end{equation}
on the polymeric phase space which transforms effective Hamiltonian
(\ref{PHamiltonian}) to the nondeformed one: $H_{\lambda}(q,p)
\rightarrow\,H_{\lambda}(q',p')$, with
\begin{equation}\label{NC-Hamiltonian}
H_{\lambda}(q',p')=\frac{p'^2}{2m}+U(q').
\end{equation}
Although Hamiltonian (\ref{NC-Hamiltonian}) is independent of the
parameter $\lambda$, the new momentum $p'$ is bounded as
$p'\in[-\frac{2}{ \lambda},+\frac{2}{\lambda})$ due to the
transformation (\ref{NC-Transformation}). Therefore, the Hamiltonian
(\ref{NC-Hamiltonian}) should be counted distinct from the standard
nondeformed Hamiltonian (\ref{Hamiltonian}). It is also important to
note that while the Hamiltonian gets the standard form, the
corresponding $2$-form in the noncanonical chart becomes
\begin{eqnarray}\label{NC-Structure}
\omega=\frac{dq'\wedge\,dp'}{\sqrt{1-(\lambda p'/2)^2}}.
\end{eqnarray}
Substituting the above symplectic structure and also the associated
Hamiltonian (\ref{NC-Hamiltonian}) into the relation (\ref{VF-D}),
we are led to the following solution for the Hamiltonian vector
field
\begin{eqnarray}\label{NC-VF}
{\bf x}_{_H}=\sqrt{1-(\lambda p'/2)^2}\Big(\frac{p'}{m}\frac{
\partial}{\partial q'}-\frac{\partial U}{\partial q'}\frac{
\partial}{\partial p'}\Big).
\end{eqnarray}
The integral curves of the above Hamiltonian vector field are the
polymer-modified Hamilton's equation in the noncanonical chart
\begin{eqnarray}\label{NC-HE}
\frac{dq'}{dt}=\frac{p'}{m}\,\sqrt{1-(\lambda p'/2)^2},\nonumber
\\{\frac{dp'}{dt}}=-\frac{\partial U}{\partial q'}\,\sqrt{1-(
\lambda p'/2)^2},
\end{eqnarray}
which will be reduced to the standard ones in the limit of
$\lambda\rightarrow\,0$. The two sets of modified Hamiltonian
equations of motion (\ref{Dar-PHE}) and (\ref{NC-HE}) are in
agreement with each other through the transformation
(\ref{NC-Transformation}).

The Poisson bracket between two real valued functions $F$ and $G$ in
this chart can be obtained by substituting the noncanonical
structure (\ref{NC-Structure}) and the associated vector field
(\ref{NC-VF}) into the definition (\ref{PB}) with the result
\begin{eqnarray}\label{NC-PB}
\{F,\,G\}=\sqrt{1-(\lambda p'/2)^2}\bigg(\frac{\partial F}{
\partial q'}\frac{\partial G}{\partial p'}-\frac{\partial F
}{\partial p'}\frac{\partial G}{\partial q'}\bigg).
\end{eqnarray}
With the help of this relation the Poisson bracket of the
noncanonical variables $q'$ and $p'$ can be obtained as
\begin{equation}\label{NC-PA}
\{q',\,p'\}=\sqrt{1-(\lambda p'/2)^2}\,,
\end{equation}
which reduces to its nondeformed counterpart in the continuum limit
$\lambda\rightarrow\,0$.

Since, by definition, the total volume is invariant over the
symplectic manifold, it should be the same as one in the Darboux
chart (\ref{Dar-TV}). Indeed, integrating the Liouville volume which
is nothing but the $2$-form structure for a two-dimensional
manifold, gives the total volume in the noncanonical chart as
\begin{eqnarray}\label{NC-TV}
\mbox{Vol}(\omega^1)=\int_{L}dq'\times\int_{-\frac{2}{
\lambda}}^{+\frac{2}{\lambda}}\frac{dp'}{\sqrt{1-(
\lambda p'/2)^2}}=2\pi\Big(\frac{L}{\lambda}\Big),\hspace{.5cm}
\end{eqnarray}
that coincides with relation (\ref{Dar-TV}) as expected. Therefore,
one can work in two equivalent pictures on the polymeric symplectic
manifold: {\it i}) utilizing the effective Hamiltonian
(\ref{PHamiltonian}) together with the symplectic structure
(\ref{Dar-twoform}) in the Darboux chart which leads to the
canonical Poisson algebra (\ref{Dar-PA}); {\it ii}) implementing the
noncanonical chart with symplectic structure (\ref{NC-Structure}),
Hamiltonian function (\ref{NC-Hamiltonian}) and the corresponding
noncanonical Poisson algebra (\ref{NC-PA}). The trajectories on the
polymeric phase space are the same in two charts since equation
(\ref{VF-D}) is satisfied in a chart-independent manner. However, as
we will see in the next section, working within the noncanonical
chart is more admissible from the statistical point of view.

\section{Density of States and Partition Function}

Statistical mechanics determines the relation between microphysics
and macrophysics. All of the thermodynamical properties of a given
physical system can be derived from its partition function which is
the summation over all accessible microstates of the system. The
canonical partition function for a single particle state is defined
as \cite{unit}
\begin{eqnarray}\label{QPF}
{\mathcal Z}_{1}=\sum_{\varepsilon}\exp\big[-\varepsilon/T
\big]\,,
\end{eqnarray}
where $\varepsilon$ are the single particle energy states which are
the solution of the Schr\"{o}dinger equation in the standard
representation of quantum mechanics. In fact, the microstates for
the statistical system are completely determined by the quantum
physics. In contrast to the standard Schr\"{o}dinger representation,
finding the energy eigenvalues in the polymer representation is not
an easy task due to the nonlinearity of the quantum Hamiltonian in
this picture \cite{QPR,QPR2,QPR3}. Even if one finds the
microstates' energies in the polymer picture, calculating the
partition function (\ref{QPF}) is somehow a complicated
issue\cite{Acosta}. Nevertheless, one may work with the classical
Hamiltonian together with the density of states in the semiclassical
regime. The semiclassical and the quantum statistics will be
coincided in the high temperature regime. Therefore one should be
careful that the semiclassical approximation is only applicable for
the polymerized systems with small polymer length scale through the
uncertainty principle.

Approximating the summation over the energy eigenvalues in the
relation (\ref{QPF}) by the integral over the phase space volume
yields
\begin{eqnarray}\label{Approx}
\sum_{\varepsilon}\rightarrow\frac{\mbox{Vol}(\omega^3)
}{h^3}=\frac{1}{h^3}\int_{\mathcal M}\omega^3\,,
\end{eqnarray}
where $\omega^3$ is the Liouville volume of the six-dimensional
phase space of the single particle which in turn, should be obtained
by substituting the associated $2$-form structure into the
definition (\ref{Vol-D}). Relation (\ref{Approx}) is written in a
chart-independent manner on the manifold since the total volume
$\mbox{Vol}(\omega^3)$ is invariant over the symplectic manifold
(see also \cite{LT}). Indeed, equation (\ref{Approx}) is nothing but
the Heisenberg uncertainty principle which implies a finite
fundamental element for the phase space volume of the order of
Planck constant. The total volume determines the number of
microstates of the system and it, as is guaranteed by the Liouville
theorem, should be invariant under the time evolution. Now, one may
consider relation (\ref{Approx}) in various charts over symplectic
manifolds with no worries about the invariance of the total volume
as time grows, since the Liouville theorem is satisfied in a
chart-independent manner via the relation (\ref{L-T}).

In the usual statistical mechanics, the topology of the phase space
of a single particle is ${\mathbb R}^6$ and then relation
(\ref{Approx}) in the Darboux (canonical) chart leads to the
well-known result,
\begin{eqnarray}\label{U-DOS}
\sum_{\varepsilon}\rightarrow\,\frac{1}{h^3}\int\int\int dx\,dy\,dz
\hspace{3cm}\\ \times\int_{-\infty}^{+\infty}\int_{-\infty}^{+\infty}
\int_{-\infty}^{+\infty}dp_x\,dp_y\,dp_z.\nonumber
\end{eqnarray}
With approximating the summation over the energy eigenvalues in
(\ref{QPF}) by relation (\ref{U-DOS}), one can obtain the standard
definition of the semiclassical partition function \cite{SMB}. In
the same way, the polymeric partition may be achieved .

We consider (\ref{Approx}) for the polymerized phase space in two
charts: the Darboux and noncanonical charts which we presented in
the previous section. In the Darboux chart, the density of states
takes the same form as the usual one (\ref{U-DOS}) because the
corresponding symplectic structure (\ref{Dar-twoform}) is canonical
(see the appendix), with this difference that now the momentum part
of the polymeric phase space has a compact topology $S^1$ rather
than the usual ${\mathbb R}$. Thus we have
\begin{eqnarray}\label{Dar-DOS}
\sum_{\varepsilon}\rightarrow\,\frac{1}{h^3}\int\int\int dx\,dy\,dz\hspace{3cm}\\
\times\int_{-\pi/\lambda}^{+\pi/\lambda}\int_{-\pi/\lambda}^{+\pi/\lambda}
\int_{-\pi/\lambda}^{+\pi/\lambda}dp_x\,dp_y\,dp_z.\nonumber
\end{eqnarray}
Approximating the quantum partition function (\ref{QPF}) by the
polymeric state density (\ref{Dar-DOS}) and substituting the
associated Hamiltonian (\ref{PHamiltonian}) instead of the energy
eigenvalues $\varepsilon$, gives the polymeric partition function as
\begin{widetext}
\begin{eqnarray}\label{Dar-PF-D}
{\mathcal Z}_1(\lambda;T,V)=\frac{1}{h^3}\int\int\int\exp\Big[
-\frac{U(x,y,z)}{T}\big]dx\,dy\,dz\times\prod_{i=1}^3\Bigg(
\int_{_{-\pi/\lambda}}^{^{+\pi/\lambda}}\exp\Big[-\frac{(1-
\cos(\lambda p_i))}{m\lambda^2T}\Big]\,d{p_i}\Bigg).
\end{eqnarray}
It is clear that both the Hamiltonian and the density of states are
modified in the polymer phase space in the Darboux (canonical)
chart. Rewriting (\ref{Approx}) in the noncanonical chart on the
polymeric symplectic manifold, gives the following polymeric density
of states in the noncanonical chart (see the appendix)
\begin{eqnarray}\label{DOS}
\sum_{\varepsilon}\rightarrow\,=\frac{1}{h^3}\int\int\int
dx'\,dy'\,dz'\times\int_{-2/\lambda}^{+2/\lambda}\int_{-2/
\lambda}^{+2/\lambda}\int_{-2/\lambda}^{+2/\lambda}\frac{
d{p'_x}\,d{p'_y}\,d{p'_z}}{\sqrt{\Big(1-(\frac{\lambda
p'_x}{2})^2\Big)\Big(1-(\frac{\lambda p'_y}{2})^2\Big)\Big
(1-(\frac{\lambda p'_z}{2})^2\Big)}}.
\end{eqnarray}
\end{widetext}
Therefore in the noncanonical chart, as the above equation shows,
while the Hamiltonian takes the standard form (\ref{NC-Hamiltonian})
the polymeric effects are summarized in the density of states.
Equation (\ref{DOS}) determines the number of accessible microstates
for the system and because of the bounded domain for the momenta
this number is less than when the polymeric effects are absent. A
similar result is also achieved in the other effective approaches to
quantum gravity such as generalized uncertainty principle
\cite{GUP-THR}, doubly special relativity \cite{DSR-THR} and
noncommutative phase space \cite{LT2}. As a first thermodynamical
outcome we can see that the polymeric effects cause a reduction in
the entropy of the system. This is because the entropy is directly
determined from the number of microstates. In the next section we
will explicitly investigate this fact and its following results for
a particular case in which the underlying system is a system of an
ideal gas.

Following the same steps which led us to (\ref{Dar-PF-D}), but this
time with the help of relations  (\ref{NC-Hamiltonian}), (\ref{QPF})
and (\ref{DOS}), the canonical polymeric partition function for a
single noninteracting particle becomes
\begin{eqnarray}\label{PF-D}
{\mathcal Z}_1(T,V)=\frac{V}{h^3}\prod_{i=1}^3\Bigg(
\int_{_{-2/\lambda}}^{^{+2/\lambda}}\exp\Big[-\frac{
{p'_i}^2}{2mT}\Big]\frac{d{p'_i}}{\sqrt{1-(\frac{\lambda
p'_i}{2})^2}}\Bigg)\nonumber\\=\frac{V}{\hbar^3\lambda^3}
\bigg(\exp\Big[-\frac{\lambda^{-2}}{mT}\Big]\,I_{0}\Big[
\frac{\lambda^{-2}}{mT}\Big]\bigg)^3,\hspace{.6cm}
\end{eqnarray}
where $V$ is the result of integration over the spatial part and
$I_0$ denotes the modified Bessel function of the first kind. The
expression (\ref{PF-D}) for the partition function is in an
excellent agreement with what is obtained in \cite{Acosta} by the
full quantum consideration. Now, the polymeric thermodynamics of a
physical system (an ideal gas in our model in the next section) may
be extracted by means of the above partition function.
\section{Ideal Gas}
In this section, let us consider a gaseous system consisting of $N$
noninteracting particles at temperature $T$ confined in the volume
$V$. We assume that this system obeys the Maxwell-Boltzmann
statistics. Equation (\ref{PF-D}) can be used to evaluate the
corresponding polymeric partition function as
\begin{eqnarray}\label{TPF-IG-D}
{\mathcal Z}_{N}(T,V)=\frac{1}{N!}\big[{\mathcal Z}_1
(T,V)\big]^N\,,
\end{eqnarray}in which the Gibb's factor is also considered \cite{SMB}. The natural choice for
the polymer length $l_{\rm poly}$ is the Planck length $l_{\rm
poly}=\alpha\hbar\lambda=\alpha\,l_{_{\rm Pl}}= \sqrt{G\hbar}$,
where $G$ is the gravitational constant and $\alpha$ is a
dimensionless coefficient of the order of unity $\alpha\sim
{\mathcal O}(1)$. As we have mentioned before, this coefficient
should be fixed only by the experiments \cite{QGExperiment}. In our
study, the value of the coefficient $\alpha$ determines the boundary
in which the polymeric effects become important. As we will see, the
polymeric effects appear in the trans-Planckian regime for the
values $\alpha<1$ and the sub-Planckian polymeric effects emerge for
$\alpha>1$. Here, we assume $l_{\rm poly}=l_{_{\rm Pl}}$, i.e, we
select $\alpha=1$ for the sake of simplicity. Substituting relation
(\ref{PF-D}) into (\ref{TPF-IG-D}) gives the total partition
function for the polymerized ideal gas as
\begin{eqnarray}\label{TPF-IG}
{\mathcal Z}_{N}(T,V)=\frac{\big(V/l_{_{\rm Pl}}^3\big)^N
}{N!}\Bigg(\exp\bigg[-\frac{T_{_{\rm Pl}}^2}{mT}\bigg]I_0
\bigg[\frac{T_{_{\rm Pl}}^2}{mT}\bigg]\Bigg)^{3N},
\hspace{.7cm}
\end{eqnarray}
where $T_{_{\rm Pl}}$ is the Planck temperature $T_{_{\rm
Pl}}=\sqrt{\frac{\hbar}{G}}$. The prefactor $\big(V/l_{_{\rm
Pl}}^3\big)\in {\mathbb N}$ in this equation shows the discreteness
of space in the polymer framework. In the following, by means of
this expression for the total partition function, we will
investigate the thermodynamical properties of the polymeric ideal
gas.

\subsection{Pressure}
First of all, let us look at the Helmholtz free energy $F$ for the
polymeric ideal gas which can be obtained (with the help of
(\ref{TPF-IG})) from its standard definition as
\begin{eqnarray}\label{Helmholtz}
F=-T\ln\big[{\mathcal Z}_{N}(T,V)\big]\hspace{3.5cm}\nonumber\\
=NT\Bigg(\ln\Bigg[\frac{N {l_{_{\rm Pl}}^3}}{VI^3_0\big[
T_{_{\rm Pl}}^2/mT\big]}\Bigg]-1\Bigg)+\frac{3NT_{_{\rm Pl}}^2}{m},
\end{eqnarray}
in which we have used the Stirling's approximation $\ln[N!]\approx
N\ln[N]-N$ for large $N$. The pressure by definition is
\begin{eqnarray}\label{pressure}
P=-\bigg(\frac{\partial F}{\partial V}\bigg)_{T,N}=\frac{NT}{V}\,.
\end{eqnarray}
Therefore, the familiar equation of state for the ideal gasses, that
is,
\begin{equation}\label{EOS}
PV=NT,
\end{equation}
preserves its form also in the polymer framework.
\subsection{Internal energy}
\begin{figure}
\flushleft\leftskip+3em{\includegraphics[width=2.5in]{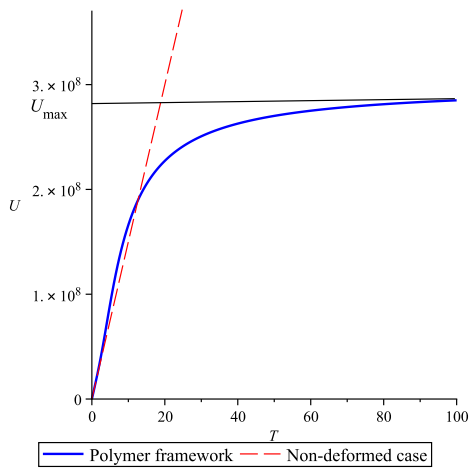}}
\hspace{3cm}\caption{\label{fig:1} The internal energy versus the
temperature; the solid line represents the internal energy in the
polymer framework and the dashed line corresponds to the standard
nondeformed case. Clearly, there is an upper bound for the internal
energy in the polymer framework which is originated in the existence
of the maximal momentum (UV cutoff) in the polymeric systems. The
figure is plotted in unit $\hbar=1$ and $T_{_{\rm Pl} }=10$, where
$T_{_{\rm P}}$ is the Planck temperature. The number of particles
and the mass are taken as $N=10^7$ and $m=10$. The polymeric effects
dominate when the temperature approaches the Planck temperature.}
\end{figure}

The internal energy of the polymeric ideal gas will be
\begin{eqnarray}\label{energy}
U=-T^2\bigg[\frac{\partial}{\partial T}\bigg(\frac{F}{T}\bigg)
\bigg]_{N,V}=\frac{3NT_{_{\rm Pl}}^2}{m}\Bigg(1-\frac{I_1\big[
T_{_{\rm Pl}}^2/mT\big]}{I_0\big[T_{_{\rm Pl}}^2/mT\big]}\Bigg)
.\hspace{.6cm}
\end{eqnarray}
This result exactly coincides with what comes from the full quantum
consideration of the ideal gas in the polymer picture but now in a
simpler manner \cite{Acosta}. It is important to note that we are
working in the semiclassical regime while our results are in good
agreement with their full quantum counterparts in the limit of high
temperature. Therefore, the quantum considerations preserve their
importance for the low temperature phenomenons such as Bose-Einstein
condensation. Nevertheless, considering the low temperature behavior
is useful even in the semiclassical regime to see how the results in
this limit may be recovered. To do this, let us take the low
temperature limit of the relation (\ref{energy}), that is
\begin{eqnarray}\label{energy-LT}
U\approx\,U_0\bigg(1+\frac{mT}{4T_{_{\rm Pl}}^2}\bigg),
\end{eqnarray}
where $U_0=\frac{3}{2}NT$ is the well-known usual internal energy
for the ideal gas. A glance at \ref{energy-LT}) shows that while the
polymeric effects becomes important at the high temperatures, the
standard relation for the internal energy of an ideal gas is
recovered in the limit of low temperature. The usual internal energy
(dashed line) and its polymeric counterpart (solid line) versus the
temperature are plotted in figure \ref{fig:1}. As this figure shows,
the two curves coincide at low temperatures and while the
temperature rises are separated from each other.

To estimate the order of magnitude of the polymeric correction to the
internal energy of an ideal gas, consider
\begin{equation}\label{Estimation}
\Big|\frac{\Delta U}{U}\Big|=\Big|\frac{U-U_0}{U}\Big|\sim\Big(\frac{m}{
m_{_{\rm Pl}}}\Big)\times\Big(\frac{T}{T_{_{\rm Pl}}}\Big),
\end{equation}
where $m_{_{\rm Pl}}$ is the Planck mass (equal to the Planck
temperature in the units in which we are working). Since we have set
numerical factor $\alpha$ to be of the order of unity, the polymeric
effects become important on the trans-Planckian regime. In this
regard, these effects will be very small in the currently accessible
temperatures \cite{expriment}. For instance, consider an ideal gas
consisting of electrons with mass $m_e\approx0.5\,{\rm MeV}$. For
the temperature about $T\sim\,1\,{\rm TeV}$ the polymeric correction
to the corresponding internal energy is of the order of
\begin{eqnarray}
\Big|\frac{\Delta U}{U}\Big|\sim\,10^{-38}\,,
\end{eqnarray}
where we have set $m_{_{\rm Pl}}\approx1.2\times\,10^{19}\,{\rm
GeV}$.

Furthermore, the usual internal energy for an ideal gas $U_0$ is linearly
proportional to its temperature and consequently the system can have
access to any arbitrary high energy scale just by sufficiently increasing
its temperature. However, a glance at the corresponding relation for
the polymeric ideal gas given by equation (\ref{energy}) shows the
existence of a finite maximum value in the high temperature limit as
(see figure \ref{fig:1})
\begin{eqnarray}\label{max-energy}
U\leq U_{\rm max}=\frac{3NT_{_{\rm Pl}}^2}{m}\,.
\end{eqnarray}
The fact that no energy scale is accessible greater than the above
upper bound may be attributed to fact that the momenta of the
noninteracting particles of the gaseous system are bounded in
polymer framework.
\subsection{Entropy}
\begin{figure}
\flushleft\leftskip+3em{\includegraphics[width=2.5in]{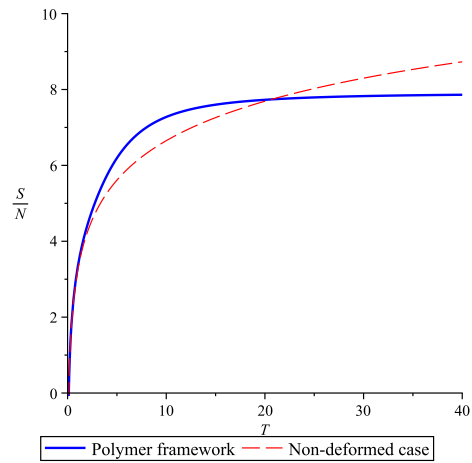}}
\hspace{3cm}\caption{\label{fig:2} The entropy against the
temperature; the solid line represents the entropy in the polymer
picture and the dashed line corresponds to the nondeformed case. The
entropy increases with a much less rate in comparison with the
ordinary ideal gas since the number of microstates in the polymer
picture is less than the usual case. The figure is plotted for
$V=10^7$.}
\end{figure}

Now let us see how the entropy changes its form under the framework
we are dealing with. A straightforward calculation based on the
Helmholtz energy (\ref{Helmholtz}) will arrive us to
\begin{eqnarray}\label{entropy}
S=-\bigg(\frac{\partial F}{\partial T}\bigg)_{N,V}=N\Bigg(\ln
\Bigg[\frac{VI^3_0\big[T_{_{\rm Pl}}^2/mT\big]}{N {l_{_{\rm Pl}}^3}}
\Bigg]+1\Bigg)\nonumber\\-\frac{3NT_{_{\rm Pl}}^2}{mT}\frac{I_1\big[
T_{_{\rm Pl}}^2/mT\big]}{I_0\big[T_{_{\rm Pl}}^2/mT\big]}.
\hspace{.3cm}
\end{eqnarray}
In figure \ref{fig:2}, we have plotted the behavior of entropy in
terms of temperature. As this figure shows entropy increases with a
much less rate in comparison with the usual ideal gas. This result
is due to the fact that entropy is directly related to the number of
microstates of the system and this quantity in turn reduces in the
polymer framework since there is an upper bound for the momentum of
the particles in this picture.

To see how the entropy behaves in the low temperature regime, we may
take this limit of the relation (\ref{entropy}) with result
\begin{eqnarray}\label{entropy-LT}
S\approx\,N\Bigg(\ln\Bigg[\frac{V}{N}\Big(\frac{2\pi mT}{h^2}
\Big)^{3/2}\Bigg]+\frac{5}{2}\Bigg)-\frac{3}{4}N\frac{mT}{
T_{_{\rm Pl}}^2}.
\end{eqnarray}
The last term on the right-hand side is the first order polymeric
correction to the entropy of an ideal gas which is negligible in the
limit of low temperatures. This is also clear from the figure
\ref{fig:2}, which shows that the entropy curve (\ref{entropy})
coincides with its standard nondeformed one in the limit of low
temperature. The order of magnitude of the polymeric correction to
the entropy is the same as we obtained for the internal energy.

\subsection{Specific heat}
\begin{figure}
\flushleft\leftskip+3em{\includegraphics[width=2.5in]{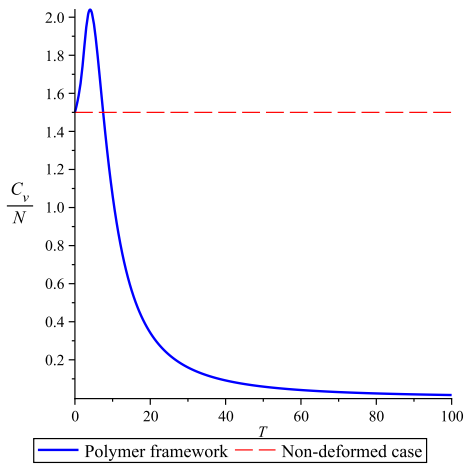}}
\hspace{3cm}\caption{\label{fig:3} The specific heat versus the
temperature in the polymer framework. Since the increasing behavior
of the internal energy stops after reaching to its maximum, it is
expected that the specific heat goes to zero after reaching to a
maximum value. The figure clearly shows this behavior.}
\end{figure}

Finally, another important thermodynamical quantity is the specific
heat which can be evaluated from the internal energy (\ref{energy})
as follows
\begin{eqnarray}\label{SHeat}
C_{_V}=\bigg(\frac{\partial U}{\partial T}\bigg)_{V}\hspace{5.2cm}
\nonumber\\=\frac{3NT_{_{\rm Pl}}^4}{m^2T^2}\Bigg(1-\frac{mT}{
T_{_{\rm P}}^2}\frac{I_1\big[T_{_{\rm Pl}}^2/mT\big]}{I_0\big[
T_{_{\rm Pl}}^2/mT\big]}-\frac{I_1^2\big[T_{_{\rm Pl}}^2/mT\big]}{
I_0^2\big[T_{_{\rm Pl}}^2/mT\big]}\Bigg).
\end{eqnarray}The specific heat versus the temperature is shown in figure
\ref{fig:3}. We see that this quantity grows until it reaches a
maximum value and then takes a decreasing behavior, eventually tends
to zero. This means that from now on, if the system gets more
thermal energy, its internal energy does not change. Such a behavior
is expected because we have seen from (\ref{max-energy}) that the
system eventually reaches a saturated internal energy. Again, our
result coincides with one arises from the full quantum consideration
of the ideal gas for the small polymer length \cite{Acosta}.

As we have done in the last two subsections, let us take a look at
the low temperature limit of the specific heat, that is
\begin{eqnarray}\label{SHeat-LT}
C_{_V}\approx\,\frac{3}{2}N\bigg(1+\frac{mT}{2T_{_{\rm
Pl}}^2}\bigg)\,.
\end{eqnarray}
Again, it is seen that the polymeric correction is of the order of
one that is obtained for the internal energy. Also, as we can see
from (\ref{SHeat-LT}), the specific heat of the polymeric system is
coincided with its value for the ordinary ideal gas at very low
temperatures.

\section{Summary and Conclusions}
The polymer picture of quantum mechanics is an exotic representation
of the commutation relations which have been investigated in a
symmetric sector of loop quantum gravity. We argued that in order to
study the statistical mechanics of a polymeric systems, analytical
evaluation of the energy eigenvalues may be impossible since the
Hamiltonian gets a nonlinear form in this framework. On the other
hand, we showed one can work with the classical Hamiltonian and the
density of states in the semiclassical regime. The advantage of this
method is that there is no need to solve the nonlinear eigenvalue
problem. Therefore, we considered the symplectic structure of the
polymeric phase space and for studying the statistical mechanics in
this picture we used the effective Hamiltonian (\ref{PHamiltonian})
and the deformed density of states (\ref{Dar-DOS}) in the Darboux
chart. Moreover, we introduced a noncanonical chart on the polymeric
symplectic manifold in which the Hamiltonian takes the standard form
(\ref{NC-Hamiltonian}) and all the polymeric effects were summarized
in the deformed density of states (\ref{DOS}). We explained that
working in this chart is more admissible from the statistical point
of view. This is because the density of states determines the number
of accessible microstates of the system and consequently the
polymeric effects only change this quantity in this chart. According
to our calculations the number of microstates decreased when the
polymeric considerations came into the play and we linked this
phenomena to the fact that the momenta are bounded in such a
framework. Based on the deformed density of states, we obtained the
canonical partition function for the polymeric many particle systems
and then utilized it to study thermodynamics of the ideal gas. In
this regard, some thermodynamical quantities of the polymeric ideal
gas such as pressure, internal energy, entropy, and specific heat
are evaluated. Having the same form of the equation of state as the
ordinary ideal gas, existence of an upper limit for the internal
energy unlike the conventional case in which there is no restriction
for the system to achieve any level of energy, increasing behavior
of the entropy but with a rate much less than the usual case and
tending to zero after reaching to a maximum value for the specific
heat were the main features of our analysis based on the ideas we
have designed in this article. As an estimation, our calculations
predict $\sim\,10^{-38}$ (in the ${\rm TeV}$ temperature scale) for
the order of magnitude of the polymeric corrections to the
thermodynamical quantities of an ideal gas. We saw that these
results are in very good agreement with their counterparts arisen
from the full quantum consideration at high temperatures and they
coincide with their usual counterparts at the low temperature limit,
so this may be considered as evidence that the way we have moved in,
was a right way.

\appendix
\renewcommand{\theequation}{A-\arabic{equation}}
\setcounter{equation}{0}
\section{Polymeric Density of States}
To analyze the thermodynamics of the polymeric ideal gas, we need to
consider a six-dimensional phase space corresponding to the single
particle state. The homogenous polymerization for the
six-dimensional phase space with variables $q^i=(x,y,z)$ and
$p_i=(p_x,p_y,p_z)$ is read from (\ref{Polymerization}) as
\begin{eqnarray}\label{3-PT}
{\mathcal{P}}[F(q^i)]=F(q^i),\hspace{1.5cm}{\mathcal{P}}[p_i]=\frac{
1}{\lambda}\,\sin(\lambda p_i),\nonumber\\{\mathcal{P}}[p_{i}^2]=
\frac{2}{\lambda^2}(1-\cos(\lambda p_i)).\hspace{1.5cm}
\end{eqnarray}
In the Darboux chart, the canonical structure for the
six-dimensional symplectic manifold is
\begin{eqnarray}\label{3CSS}
\omega=dx\wedge dp_x+dy\wedge dp_y+dz\wedge dp_z,
\end{eqnarray}
where the momenta are bounded as
$p_i\in[-\frac{\pi}{\lambda},+\frac{\pi}{\lambda})$. The associated
$6$-form Liouville volume can be obtained by substituting (\ref{3CSS})
into the definition (\ref{Vol-D}) as
\begin{eqnarray}\label{L-3CSS}
\omega^3=dx\wedge dy\wedge dz\wedge dp_x\wedge dp_y\wedge dp_z.
\end{eqnarray}
The density of states corresponding to the structure (\ref{3CSS})
can be obtained via the relation (\ref{Approx}) as
\begin{eqnarray}\label{3DOS-Dar}
\sum_{\varepsilon}\rightarrow\frac{1}{h^3}\int
d^3q\times\int_{|p_i|<\frac{\pi}{\lambda}}d^3p,
\end{eqnarray}
which clearly gives the relation (\ref{Dar-DOS}). To obtain the
state density in the noncanonical chart, we apply transformation
(\ref{NC-Transformation}) to the canonical variables $(q^i,p_i)$ in
the polymeric six-dimensional phase space as
\begin{equation}\label{3NC-Transformation}
(q^i,p_i)\rightarrow\,(q'^i,p'_i)=\Big(q^i,\frac{2}{
\lambda}\sin(\frac{\lambda p_i}{2})\Big),
\end{equation}
where $q'^i=(x',y',z')$ and $p'_i=(p'_x,p'_y,p'_z)$. The $2$-form
symplectic structure for the six-dimensional symplectic manifold in
the noncanonical chart becomes
\begin{eqnarray}
\omega=\frac{dx'\wedge dp'_x}{\sqrt{1-(\frac{\lambda p'_x}{2})^2}}
+\frac{dy'\wedge dp'_y}{\sqrt{1-(\frac{\lambda p'_y}{2})^2}}+
\frac{dz'\wedge dp'_z}{\sqrt{1-(\frac{\lambda p'_z}{2})^2}},
\hspace{.7cm}
\end{eqnarray}
which leads to the $6$-form Liouville volume
\begin{eqnarray}\label{L6D}
\omega^3=\frac{dx'\wedge dy'\wedge dz'\wedge dp'_x\wedge dp'_y\wedge dp_z'}{
\sqrt{\Big(1-(\frac{\lambda p'_x}{2})^2\Big)\Big(1-(\frac{
\lambda p'_y}{2})^2\Big)\Big(1-(\frac{\lambda p'_z}{2})^2\Big)}},
\hspace{.7cm}
\end{eqnarray}
through the definition (\ref{Vol-D}). The density of states (\ref{DOS})
can be easily deduced by substituting the Liouville volume (\ref{L6D})
into the relation (\ref{Approx}). Relation (\ref{DOS}) is essential to
obtain the partition function in polymer framework.


\begin{thebibliography}{1}
\bibitem{PS-Topology} In momentum polarization,
it becomes the Bohr compactification of the real
line ${\mathbb R}_{_B}$ \cite{QPR2} that leads
to the compact topology for the momentum part of
the corresponding phase space (see also
\cite{LT}).
\bibitem{unit} We work in unit $k_{_B}=1=c$,
where $k_{_B}$ and $c$ are the Boltzman
constant and speed of light in vacuum
respectively.
\bibitem{expriment} Here it is important to note
that the correction term is also closely related to the free
numerical parameter $\alpha$ which we have set to be of order of
unity. In fact the numerical factor $\alpha$ determines the
fundamental length scale of the quantum gravity (or lattice
discreteness length) as $l_{_{\rm poly}}=\alpha\,l_{_{\rm Pl}} $.
Preserving this parameter, one can find an upper bound for the the
polymer length scale $l_{_{\rm poly}}$ (see for instance
\cite{Acosta2}).
\end{thebibliography}
\end{document}